\documentclass[11pt,a4paper]{article}
\usepackage{amsmath,amssymb,amsfonts}
\usepackage{graphicx,a4wide}
\usepackage[top=3.0cm,bottom=2cm,left=1.5cm,right=1.5cm]{geometry}

\newcommand{\be}{\begin{equation}}
\newcommand{\ee}{\end{equation}}
\newcommand{\bea}{\begin{eqnarray}}
\newcommand{\eea}{\end{eqnarray}}
\newcommand{\nn}{\nonumber}

\newcommand{\al}{\alpha}
\newcommand{\gm}{\gamma}
\newcommand{\Gm}{\Gamma}
\newcommand{\ep}{\epsilon}
\newcommand{\de}{\delta}
\newcommand{\De}{\Delta}
\newcommand{\om}{\omega}

\newcommand{\lm}{\lambda}
\newcommand{\Lm}{\Lambda}
\newcommand{\sg}{\sigma}

\newcommand{\ov}{\overline}

\newcommand{\oq}{\overline{q}^{\,2}}

\newcommand{\op}{\overline\Pi}
\newcommand{\iop}{{\rm Im}\,\overline\Pi}
\newcommand{\og}{\overline G}  
\newcommand{\wt}{\widetilde}

\newcommand{\vk}{\vec k}
\newcommand{\vp}{\vec p}
\newcommand{\vq}{\vec q}

\newcommand{\la}{\langle}
\newcommand{\ra}{\rangle}

\newcommand{\ps}{p \!\!\! /}
\newcommand{\ks}{k \!\!\! /}

\newcommand{\mn}{{\mu\nu}}
\newcommand{\del}{\partial}

\newcommand{\F}{F_\pi}
\newcommand{\opq}{\vp\cdot\vq}

\begin{document}

\setcounter{page}{1}

\title{$\rho$ self energy
at finite temperature and density in the real-time formalism}
\author{Sabyasachi Ghosh and Sourav Sarkar}
\maketitle
\begin{center}
\it{Theoretical Physics Division, Variable Energy Cyclotron Centre,\\
1/AF, Bidhannagar, 
Kolkata 700064, India}
\end{center}

\begin{abstract}
 
The $\rho$ meson self-energy in nuclear matter from baryonic loops
is analysed in the real time formulation
of field theory at finite temperature and density. The discontinuities
across the branch cuts of the self-energy function are
evaluated for an  exhaustive set of resonances in the loops considering the
fully relativistic thermal baryon propagator including anti-baryons. Numerical calculations 
show a significant broadening of the $\rho$ spectral function coming from the Landau
cut. Adding the contribution from mesonic loops, the full spectral function of the $\rho$
in a thermal gas of mesons, baryons and antibaryons in equilibrium 
is evaluated at various values of temperature and baryonic
chemical potential.

\end{abstract}

%\pacs{11.10.Wx}
%\maketitle
%\keyword
\section{Introduction}

It is well known that the rate of dilepton production from a thermal
system is proportional to the two-point correlation function of vector
currents.
Hence the spectral function of vector meson, the $\rho$ meson particular,
plays such an important role in the analysis of the late stages of
heavy ion collisions~\cite{Annals,RappAdv}. The NA60 experiment at the CERN SPS measured dimuon pairs in
In-In collisions in which an excess was observed over the contribution 
from hadronic decays at freeze-out in the mass region below the $\rho$ 
peak~\cite{na60}. This was attributed to the broadening of the $\rho$ 
in hot and dense medium~\cite{RappAdv}. More recently, the PHENIX experiment 
reported a substantial excess of electron pairs in the same region of
invariant mass~\cite{Phenixdil}. This has been investigated by several
groups but the yield in all these cases have remained
insufficient to explain the data. Thus the issue of low mass 
lepton pair yield in heavy ion collisions is far from closed
and is one of the key issues to be addressed in the forthcoming
 Compressed Baryonic Matter(CBM)
experiment to be performed at the FAIR facility in GSI~\cite{CBM}.

A substantial volume of work has been devoted to the study
of $\rho$ meson properties in hot and dense medium.
We do not attempt to review the existing literature but mention a few of them
to put our work in perspective.
We find that it is only for the $\pi-\pi$ loop~\cite{Gale,Friman}
that one calculates the thermal 
loop directly. In the case of other loops
typically involving one heavy and one light particle or both heavy particles 
one uses in general either the virial formula~\cite{Smilga,Post1,Eletsky}
or the Lindhard function~\cite{Fetter,RappNPA,Peters,Dani,Post2}. 
Most of the calculations involving baryonic effects mentioned above
were performed at zero temperature. Finite temperature 
effects on the $\rho$ spectral function in dense matter have been evaluated by 
Rapp et al~\cite{RappNPA} in terms of resonant interactions of the $\rho$ with
surrounding mesons and baryons in addition to modifying the pion cloud.
Eletsky~\cite{Eletsky} and collaborators have also evaluated 
the spectral function of vector mesons at finite temperature and density
in terms of forward scattering amplitudes 
constructed  using experimental inputs assuming resonance dominance at low
energies and a Regge-type approach at higher energies.

The sources modifying the free propagation
of a particle find a unified description in terms of contributions from the 
branch cuts of the self energy function as shown 
by Weldon~\cite{Weldon}. In addition to the unitary cut present 
already in vacuum, the thermal amplitude generates a new
cut, the so called the Landau cut which provides the effect of collisions 
with the surrounding particles in the medium. This formalism was applied 
to obtain the $\rho$ self-energy in hot mesonic matter~\cite{Ghosh1} by evaluating the
one loop self-energies involving the $\pi$, $\omega$, $h_1$  and  $a_1$ 
mesons. A significant broadening of the spectral function was obtained without
appreciable shift in the mass as expected from chiral interactions.

In this work, we extend this analysis to the case of baryonic matter at
finite temperature considering an exhaustive set of 4-star resonances in the 
baryonic loops making up the $\rho$ self-energy. 
The framework of real time 
thermal field theory~\cite{Umezawa,Niemi,Kobes,Mallik_RT} that we use, enables us to evaluate 
the imaginary part 
of the self-energy from the branch cuts for real and positive values of
energy and momentum without having to resort to analytic continuation
as in the imaginary time approach~\cite{Matsubara}.
Here we work with the full relativistic baryon propagator
in which baryons and anti-baryons manifestly appear on an equal footing. Thus
the contributions from all the 
singularities in the self-energy function including the distant ones coming from 
the unitary cut
of the loops involving heavy baryons are also included. 
These are usually not considered but can contribute appreciably to the real part
of the $\rho$ meson self-energy as shown~\cite{Ghosh3} in the case of a $N\Delta$
loop. In addition
we have used
the covariant form of the momentum dependent vertex functions in the loop 
integrals in which
additional terms~\cite{Peccei} required to
describe the coupling of
off-shell spin 3/2 fields have been introduced.

In the following section we define the correlation function of vector currents
and its relation to the transverse propagator of the $\rho$. We also provide the
various Lagrangian densities which will be used at the vertices of the loop
graphs. Next, in section 3 we specify the kinematic decomposition of the
thermal propagators. In section 4 we evaluate the baryonic self-energy graphs 
as well as the discontinuities across the branch cuts. Section 5 contains the results
of the numerical evaluation of the real and imaginary parts as well as the
$\rho$ spectral function followed by a summary and discussions in section 6. In
the appendix  we provide the details of various factors appearing in the expression
for self-energy for the different loops and provide some details of
evaluation of the imaginary part in addition to a brief discussion on propagators
and self-energies in the real time formalism.    

\section{The two point function in the medium}
We begin our discussion with the two point function of vector currents in vacuum,
\be
T^{ij}_{\mn}(E,\vq)=i\int d^3xd\tau\,e^{iq\cdot x}\la 0|
T V^i_\mu(x) V^j_\nu(0)|0\ra
\ee
where $V^i_\mu(x)$ are the vector currents of two flavour QCD, given by
\be
V_\mu^i(x)=\bar q(x)\gm_\mu\frac{\tau^i}{2}q(x),~~~~~~ q=\left(\begin{array}{c}
u \\ d \end{array}\right) 
\ee
$\tau^i$ being the Pauli matrices.
In the real time formulation of thermal field theory, 
the in medium two point function assumes a $2\times2$ matrix structure~\cite{Kobes}.
The thermal two point function is given by
\be
T^{ij,ab}_{\mn}(E,\vq)=i\int d^3xd\tau\,e^{iq\cdot x}\la
 T_c V^i_\mu(x) V^j_\nu(0)\ra^{ab}
\label{Tmed}
\ee
where $\la{\cal{O}}\ra$ denotes the ensemble average of an operator ${\cal{O}}$,
\be
\la{\cal{O}}\ra=Tr(e^{-\beta H}{\cal O})/Tr e^{-\beta H}
\ee
and $Tr$ indicating trace over a complete set of states.
The superscripts 
$a,b\,(=1,2)$ are thermal indices and $T_c$ denotes time ordering with respect
to a contour in the plane of the complex time variable~\cite{Mallik_RT}.
The two point function of vector currents can be related to the $\rho$ meson
propagator using the method of external fields~\cite{Gasser} 
where one introduces a classical vector field $v^i_\mu(x)$
coupled to the vector current $V^i_\mu(x)$. The free propagator of the rho 
meson can be obtained by coupling the external field to the 
$\rho$ meson field operator using the Lagrangian~\cite{Mallik_VA} 
\be
{\cal L}_{\rho v}=\frac{F_\rho}{m_\rho}
\del^\mu\vec v^\nu\cdot(\del_\mu\vec\rho_\nu-\del_\nu\vec\rho_\mu)\nonumber
\ee
where F$_\rho = 154$ MeV is obtained from the decay 
$\rho^0\to e^+\,e^-$.

The transverse $\rho$ meson propagator $G^{ab}_{\mn}$ 
is then obtained from the relation $T^{ab}_{\mn}=K_\rho G^{ab}_{\mn}$ where the factor 
$K_\rho=({F_\rho q^2}/{m_\rho})^2$ 
comes from the coupling of the current with the $\rho$ field~\cite{Ghosh1}.
 The isospin structure is given by 
$\delta^{ij}$ which we omit from now on.  

\begin{figure}
\includegraphics[scale=0.8]{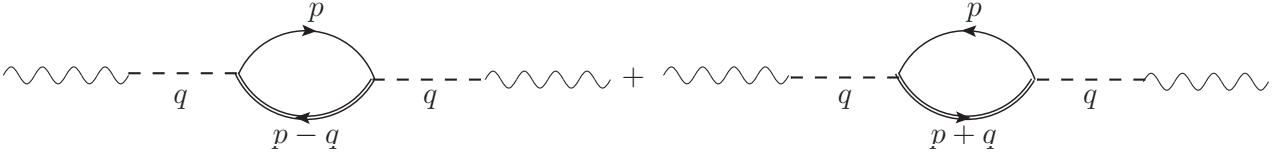}
\caption{One-loop Feynman diagrams for the two-point function contributing to
the $\rho$ self-energy in baryonic matter. The solid and double
lines stand for nucleons and resonances respectively.}
\label{loop_BB}
\end{figure}

The free propagation of the $\rho$ meson is modified by interactions in the medium which is 
populated by mesons and baryons. 
Here we consider one loop graphs shown in Fig.~\ref{loop_BB}
consisting of the nucleon $N$ and another  baryon $R$ denoted by the
double lines. 
We have included all spin one-half and three-half $4-$star
resonances listed by the PDG~\cite{PDG} so that $R$ stands for the 
$N^*(1520)$,  $N^*(1650)$, $N^*(1700)$, $\Delta(1230)$, 
$\Delta^*(1620),$ $\Delta^*(1720)$ as well as the $N(940)$ itself.
Omitting isospin factors,
the $\rho N$ couplings with the resonances 
are described by the gauge invariant interactions~\cite{Post2}    
\bea
{\cal L}&=&\frac{f}{m_\rho}[\ov{\psi}_R\sigma^{\mn}{\rho}_{\mn}\psi_N+ h.c.]
~~~~~~~~~~J^{P}_R=\frac{1}{2}^+ \nonumber\\
{\cal L}&=&\frac{f}{m_\rho}[\ov{\psi}_R\sigma^{\mn}\gamma^5{\rho}_{\mn}\psi_N+ h.c.]
~~~~~~~J^{P}_R=\frac{1}{2}^- \nonumber\\
{\cal L}&=&\frac{f}{m_\rho}[\ov{\psi}^{\mu}_R\gamma^{\nu}\gamma^5\rho_{\mn}\psi_N  + h.c.]
~~~~~~~~J^{P}_R=\frac{3}{2}^+ \nonumber\\
{\cal L}&=&\frac{f}{m_\rho}[\ov{\psi}^{\mu}_R\gamma^{\nu}\rho_{\mn}\psi_N  + h.c.]
~~~~~~~~~~~J^{P}_R=\frac{3}{2}^-
\label{lag1}
\eea
where $\rho_{\mn}=\del_\mu\rho_\nu-\del_\nu\rho_\mu$ and
$\sg^{\mn}=\frac{i}{2}[\gm^\mu\gm^\nu-\gm^\nu\gm^\mu]$.
The isospin part of the $RN\rho$ interaction is given by
\bea
&&\ov\psi_{R\, a}(\vec\tau\cdot\rho)^a_b\psi_N^b~~~~~~~~~~~~~~~~~I=1/2\nonumber\\
&&\frac{1}{\sqrt{2}}\ov\psi_{R\, abc}(\vec\tau\cdot\rho)^b_d\psi_N^a\ep^{cd}~~~~~~~I=3/2
\eea
where the indices $a,b,c,d$ take values 1 and 2 and $\ep^{12}=-\ep^{21}=1$.
For the self-energy diagrams shown in Fig 1, the isospin factor  
$I_F$ comes out to be 2 for $I=\frac{1}{2}$ and $\frac{4}{3}$ for 
$I=\frac{3}{2}$.

It is essential to point out that for the spin ${3}/{2}$ resonances 
this coupling is not quite correct
 owing to the fact that the free Lagrangian
 for the Rarita-Schwinger field $\psi^{\mu}_R$ has a free parameter~\cite{Rarita}. 
 A symmetry is associated with a point transformation under
which the free Lagrangian remains invariant up to a change in the value of
 the parameter~\cite{Peccei}. The standard practice 
is to make a choice of the value of this parameter so that the spin-3/2 
 propagator has a simple form.
In order that the interaction also remains invariant under this transformation
 an additional term is added to it.
Thus the Lagrangians involving spin-3/2 fields take the form
\bea
{\cal L}&=&\frac{f}{m_\rho}[\ov{\psi}^{\alpha}_R {\cal O}_{\alpha\beta}
\gamma_{\nu}\gm^5\rho^{\beta\nu}\psi_N  + h.c.]
~~~~~~~J^{P}_R=\frac{3}{2}^+ \nonumber\\
{\cal L}&=&\frac{f}{m_\rho}[\ov{\psi}^{\alpha}_R {\cal O}_{\alpha\beta}
\gamma_{\nu}\rho^{\beta\nu}\psi_N  + h.c.]
~~~~~~~~~~J^{P}_R=\frac{3}{2}^- 
\label{lag2}
\eea
with ${\cal O}_{\mu\alpha}=g_{\mu\alpha}-\frac{1}{4}\gm_\mu\gm_\alpha$, the
second term contributing only when the spin $3/2$ field is off the mass shell.
The value of the coupling strength $f$ thus remains unaffected by this exercise.

\section{Kinematics of the $\rho$ propagator}
The complete propagator of the $\rho$ is obtained from the Dyson
equation~\cite{Kobes,Bellac}
\be
G_{\mn}^{ab}(q)=G_{\mn}^{(0)ab}(q)- G_{\mu\lm}^{(0)ac}(q)\Pi^{\lm\sg, cd}_{\rm tot}(q)
 G_{\sg\nu}^{db}(q)
\label{dyson_G}
\ee
where $\Pi_{\rm tot}^{\mn,ab}$ denotes the thermal self-energy matrix and  
$G_{\mn}^{(0)ab}(q)$ stands for the free thermal propagator. 

As described briefly in the appendix, one can get rid of the thermal 
indices by diagonalisation. In terms of the diagonal 
elements (denoted by bar) which are analytic functions, 
the Dyson equation for the $\rho$ propagator reads
\be
\og_{\mn}(q)=\og_{\mn}^{(0)}(q)-\og_{\mu\lm}^{(0)}(q)\op^{\lm\sg}_{\rm tot}(q)
\og_{\sg\nu}(q)~,
\label{dyson_G_2}
\ee
where
\be
\og^{(0)}_{\mn}(q)=\left(-g_{\mn}+
\frac{q_\mu q_\nu}{q^2}\right)\frac{-1}{q^2-m_\rho^2+i\ep}~.
\ee
The one loop self energy  with baryons is obtained from the 
two diagrams shown in Fig.~\ref{loop_BB} so that
\be
\op^{\lm\sg}_B(q)=\op^{\lm\sg}(q)+\op^{\sg\lm}(-q)~.
\label{diag1a-b}
\ee
On addition of the contribution from the meson loops
the total $\rho$ self-energy is given by
\be
\op^{\lm\sg}_{\rm tot}(q)=\op^{\lm\sg}_B(q)+\op^{\lm\sg}_M(q)~.
\ee
In the medium, the presence of the four velocity $u_\mu$ introduces an additional 
scalar variable 
$u\cdot q$ in addition to $q^2$
leading to two independent tensors $P_{\mn}$ and $Q_{\mn}$ in terms of which 
the propagator and self-energy can be
 written as 
\bea
\og_{\mn}&=&P_{\mn}\og_t + Q_{\mn}\og_l\nonumber\\
\op_{\mn}&=&P_{\mn}\op_t + Q_{\mn}\op_l
\label{T_L}
\eea
with 
\bea
P_{\mn}&=&-g_{\mn}+\frac{q_\mu q_\nu}{q^2}-\frac{q^2}{\oq}\wt u_\mu \wt
u_\nu,~\wt u_\mu=u_\mu-(u\cdot q)q_{\mu}/q^2~;\nonumber\\
Q_{\mn}&=&\frac{(q^2)^2}{\oq}\wt u_\mu \wt
u_\nu,~\oq=(u\cdot q)^2-q^2~.
\label{defP+Q}
\eea
Using (\ref{T_L}), the Dyson equation (\ref{dyson_G_2}) can be solved to get,
\be
\og_t(q)=\frac{-1}{q^2-m_\rho^2-\op_t(q)},~~~~~
\og_l(q)=\frac{1}{q^2}\frac{-1}{q^2-m_\rho^2-q^2\op_l(q)}
\ee
where 
\be
\op_t=-\frac{1}{2}(\op_\mu^\mu +\frac{q^2}{\bar q^2}\op_{00}),~~~~
\op_l=\frac{1}{\bar q^2}\op_{00} , ~~~\op_{00}\equiv u^\mu u^\nu \op_{\mn}~.
\label{pitpil}
\ee
The self-energy function $\op_{\mn}$  can be obtained from 
the 11-component of the in-medium self-energy matrix using (see appendix)
\bea
{\rm Re}\,\op_{\mn}&=&{\rm Re}\,\Pi_{\mn}^{11}\nonumber\\
\iop_{\mn}&=&\epsilon(q_0)\tanh(\beta q_0/2){\rm Im}\,\Pi_{\mn}^{11}
\label{def_diag}
\eea
in terms of which the retarded self-energy is given by~\cite{Bellac}
\bea
{\rm Re}\,\Pi_{\mn}&=&{\rm Re}\,\op_{\mn}\nonumber\\
{\rm Im}\,\Pi_{\mn}&=&\epsilon(q_0)\iop_{\mn}~.
\label{def_ret}
\eea
We now proceed to evaluate the 11-component of the rho self-energy in the
following section.

\section{The self energy and its analytic structure}

Let us begin by writing the expression for the $\rho$ self-energy in vacuum
corresponding to the first diagram in Fig.~1.
For spin 1/2 resonances in the loop, this is given by
\be
\Pi^{\mn}(q)=i I_F\left(\frac{fF(q)}{m_\rho}\right)^2 
\int \frac{d^4p}{(2\pi)^4} Tr[\Gm^{\mu}S(p,m_N)\Gm^{\nu}S(p-q, m_R)] 
\label{spin1/2}
\ee
where $S(p,m)=(\ps+m)\Delta(p,m)$ is the fermion propagator, 
$\Delta(p,m)$ being the free propagator for a scalar field of mass $m$ and is given by
\be
\Delta(p,m)=\frac{-1}{p^2-m^2+i\ep}~.
\ee
Also included is a monopole form factor $F(q)=\Lm^2/\Lm^2+\vec q^2$ with $\Lm=2$
GeV~\cite{RappNPA} to take into account the finite size of the $\rho N R$ vertex.
%In this connection it is essential to point out that the thermal distribution
%functions present in the expressions for the self-energy already cut off 
%high momenta flowing in the loops. Thus the form
%factors are not as essential in the medium as they are for evaluation of
%graphs in vacuum. 

The corresponding expression for the case of loop graphs with spin 3/2 resonances 
is given by
\be
\Pi^{\mn}(q)=i I_F\left(\frac{fF(q)}{m_\rho}\right)^2 \int \frac{d^4p}{(2\pi)^4} 
Tr[\Gm^{\mu\alpha}S(p,m_N)\Gm^{\nu\beta}S_{\beta\alpha}(p-q, m_R)]
\label{spin3/2}
\ee 
where the spin-3/2 propagator is 
$S_{\mn}(k,m)=(\ks + m)K_{\mn}(k)( \De(k,m)$ with
$K_{\mn}(k)= -g_{\mn}+ \frac{2}{3m^2}k_\mu k_\nu + \frac{1}{3}\gamma_\mu\gamma_\nu 
+ \frac{1}{3m}(\gamma_\mu k_\nu - \gamma_\nu k_\mu )$.
Obtaining the vertex factors $\Gm^{\mu}$ and $\Gm^{\mu\alpha}$ 
from the interaction Lagrangians (\ref{lag1}) and (\ref{lag2}) 
both the expressions (\ref{spin1/2}) and (\ref{spin3/2}) can
be expressed in the general form
\be
\Pi_{\mn}(q)=i\int\frac{d^4p}{(2\pi)^4}L_{\mn}(p,q)\De (p,m_N)\De(p-q,m_R) 
\ee
where the factor $L_{\mn}(p,q)$ consists of the trace over Dirac matrices appearing 
in the two fermion propagators along with
their associated tensor structures, isospin and form factors coming
from the $\rho NR$ vertex. Since the self-energy is transverse,
$L^\mn$ can be expressed as
\be
L^{\mn}(p,q)=I_F \left(\frac{fF(q)}{m_\rho}\right)^2 [\alpha(p,q) A^{\mn}+\beta(p,q)
B^{\mn}+\gm(p,q) C^{\mn}]
\ee
where the three gauge-invariant tensors $A^{\mn}$, $B^{\mn}$ and $C^{\mn}$ are 
given by
\bea
A_{\mn}(q)&=&-g_{\mn}+{q_\mu q_\nu}/{q^2} ,\nonumber\\
B_{\mn}(q,p)&=&q^2 p_\mu p_\nu-q\cdot p(q_\mu p_\nu+p_\mu q_\nu)
+(q\cdot p)^2g_{\mn} ,\nonumber\\
C_{\mn}(q,p)&=&q^4 p_\mu p_\nu-q^2(q\cdot p)(q_\mu p_\nu+p_\mu
q_\nu) +(q\cdot p)^2q_\mu q_\nu~.
\label{ABC}
\eea
The coefficient functions $\alpha(p,q)$, $\beta(p,q)$ and $\gm(p,q)$ for the different loops are tabulated in
the appendix.

We now extend the vacuum self-energy to the nuclear medium.
In the real-time version of thermal field theory 
that we are using 
the propagators assume the form of matrices. 
The spin and isospin structure of the self-energy graph remaining the same, 
it is only the scalar part $\Delta(p,m)$ of the propagators that assumes 
a matrix structure. 
The required $11-$component of the fermion propagator is given by 
\be
E^{11}(p)=\Delta(p)+2\pi i
N(p_0)\delta(p^2-m^2);~~N(p_0)=n_+(\omega)\theta(p_0)+n_-(\omega)\theta(-p_0)~.
\label{de11}
\ee
The function $n_\pm(\omega)=\displaystyle\frac{1}{e^{\beta(\omega \mp \mu)}+1}$ is
the Fermi distribution where the $\pm$ sign in the subscript refers to baryons and 
anti-baryons respectively, $\om=\sqrt{\vp^2+m^2}$ 
and $\mu$ is the baryonic chemical potential which is 
taken to be equal for all the baryons considered here. Expressed as
\be
E^{11}(p)=-\frac{1}{2\om}\left(\frac{1-n_+}{p_0-\om+i\ep}+
\frac{n_+}{p_0-\om-i\ep}-\frac{1-n_-}{p_0+\om-i\ep}
-\frac{n_-}{p_0+\om+i\ep}\right)
\label{fullprop}
\ee
the first and the second terms can be identified with the propagation of baryons
above the Fermi sea and holes in the Fermi sea respectively~\cite{Vol_16} while
the third and
fourth terms correspond to anti-baryons. 

As noted in the previous section, the in-medium self-energy function of the
$\rho$
can be obtained from the 11-component of the thermal self-energy 
matrix. For the one-loop graphs shown in Fig.~1, the latter is given by
\be
\Pi^{11}_{\mn}(q)=i\int\frac{d^4p}{(2\pi)^4}L_{\mn}(p,q)
E^{11}(p,m_N)E^{11}(p-q,m_R) ~.
\label{eq_b}
\ee
Upon inserting the form of $E^{11}$ from (\ref{de11}) we get three types of 
terms. One is the vacuum contribution involving 
the vacuum parts of the two propagators, the other two being medium dependent, 
one linear and the other quadratic in the
thermal distribution function. 
Performing the $p_0-$integration and using the relations (\ref{def_diag})
connecting the real and imaginary parts of the $11$-component of the 
self-energy matrix with those of the diagonal 
element (defined using a bar), the self-energy function is written as
\bea
\ov\Pi^\mn(q_0,\vq)=\int\frac{d^3\vec p}{(2\pi)^3 4\om_N\om_R}&\times &
\left[\frac{L^\mn_1 n^N_+ -L^\mn_3 n^R_+}{q_0 -\omega_N+\omega_R+i\epsilon(q_0)\eta}-
\frac{L^\mn_2 n^N_- -L^\mn_4 n^R_-}{q_0 +\omega_N-\omega_R+i\epsilon(q_0)\eta}
\right. \nonumber\\
&&\left.+\frac{L^\mn_1 (1-n^N_+) -L^\mn_4 n^R_-}{q_0 -\omega_N-\omega_R+i\epsilon(q_0)\eta}
-\frac{L^\mn_2 (1-n^N_-) -L^\mn_3 n^R_+}{q_0
+\omega_N+\omega_R+i\epsilon(q_0)\eta}\right] 
\label{Pi_a}
\eea 
where $n^N\equiv n(\omega_N)$ with $\omega_N=\sqrt{\vp^2+m_{N}^2}$, $n^R\equiv 
n(\omega_R)$ with $\omega_R=\sqrt{(\vp-\vq)^2+m_{R}^2}$
and $L^\mn_i,i=1,..4$ denote the values of $L^\mn(p_0)$ for
$p_0=\om_N,-\om_N,q_0+\om_R,q_0-\om_R$ respectively.

Let us first consider the imaginary part of the self-energy. 
The retarded self-energy defined by (\ref{def_ret}) can be
easily read off from the self-energy function (\ref{Pi_a}) to get
\bea
&& {\rm Im} \Pi^\mn(q_0,\vq)=-\pi\coth(\beta q_0/2) \int\frac{d^3\vec p}{(2\pi)^3 4\om_N\om_R}\times  \nonumber\\
&&[L^\mn_1\{(1-n^N_+-n^R_-)\de(q_0-\om_N-\om_R)
+(n_+^N-n_+^R)\de(q_0-\om_N+\om_R)\}\nonumber\\
&& + L^\mn_2\{(n_-^R-n_-^N)\de(q_0+\om_N-\om_R)
-(1-n_-^N-n_+^R)\de(q_0+\om_N+\om_R)\}]
\label{ImPi_a}
\eea 
in which the factors $L^\mn_{3,4}$ have converted to $L^\mn_{1,2}$ respectively in 
association with the $\de$-functions.
Following~\cite{Weldon,Ghosh1} it is interesting to relate the
terms appearing in the above expression with scattering and decay processes 
involving the $\rho$, nucleon $(N)$ and the heavy resonances $(R)$ 
 in the medium. The delta functions in each of the terms in (\ref{ImPi_a}) 
 precisely define the kinematic domains where these processes can occur. 
 The regions which are non-vanishing 
  give rise to cuts in the self-energy function. Thus, the first and 
 the fourth terms are non-vanishing for $q^2>(m_R+m_N)^2$
 giving rise to the unitary cut and second and third terms are non-vanishing for
  $q^2<(m_R-m_N)^2$ giving rise to the Landau cut. Note that
the unitary cut is present in vacuum but the Landau cut appears
 only in the medium. 

Consider, for example the first term with $1-n^N_+-n^R_-$ in (\ref{ImPi_a}). Written as 
$(1-n^N_+)(1-n^R_-)-n^N_+n^R_-$ this indicates a process in which
 a  (virtual) $\rho$ decays into a $NR^{-1}$ pair with the 
 Pauli blocked probability  $(1-n^N_+)(1-n^R_-)$ minus the process in
 which  $NR^{-1}$ pair gets absorbed in the medium with a statistical weight
  factor $n^N_+n^R_-$. This process can obviously 
take place for $\rho$'s with invariant mass $(\sqrt{q^2}) > (m_R+m_N)$,
 a requirement that is in conformity with the kinematic threshold of the
unitary cut coming from the associated $\delta-$function. 

\begin{figure}
\centerline{\includegraphics[scale=0.6]{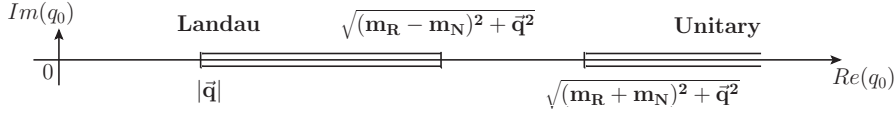}}
\caption{The positions of the Landau and unitary cuts in the complex $q_0$
plane for $q^2>0$.}
\label{cutfig}
\end{figure}  

The kinematic domains where the four terms contribute can be summarised 
as follows. For
$q_0>0$, the first term in (\ref{Pi_a}) contributes for time-like values of
$q^2$, the second at space-like $q^2$ and the third at all $q^2$. Likewise, 
for negative values of the variable $q_0$, the second term is non-zero
at all $q^2$, the third at space-like $q^2$ and the fourth only for time-like
$q^2$. In view of the fact that the spectral function of the $\rho$ will be 
measured in the invariant mass spectra of lepton pairs we will henceforth
confine ourself to the kinematic region $q_0>0$ and  $q^2>0$.
The position of the relevant cuts in the complex energy plane 
for this region are shown in Fig.~\ref{cutfig} where we have ignored the portion
of the Landau cut for $q_0<|\vq|$. (For a discussion on 
the branch cuts on the entire $q_0$ axis, see~\cite{Ghosh1}).
Furthermore, we will not include the unitary cut contribution
in the imaginary part since the 
threshold of this cut begins at $q_0>m_R+m_N$
which being far away from the $\rho$ pole is not expected to contribute 
to the $\rho$ spectral 
function in a substantial way. Thus only the Landau cut as given by the third
term in eq.~(\ref{ImPi_a}) will be considered.

Collecting the Landau contributions from both the diagrams using (\ref{diag1a-b})
we finally write down the imaginary part of the $\rho$ self-energy due to baryonic 
loops
\bea
&&{\rm Im}\Pi_B^\mn(q_0,\vq)=\pi\int\frac{d^3\vec p}{(2\pi)^3 2\om_N}
\left[\frac{L_1^\mn (-q)}{2\om_{R'}}\{(n^N_+ -n^{R'}_+)\de(q_0+\om_N-\om_{R'})\}\right.\nonumber\\
&& + \left.\frac{L_2^\mn (q)}{2\om_R}\{(n^N_- - n^R_-)\de(q_0+\om_N-\om_R)\}\right]
\eea
where $\omega_{R'}=\sqrt{(\vp+\vq)^2+m_{R}^2}$. The two terms in this
expression describe the contributions from scattering processes. The factor 
$(n_+^N-n_+^{R'})$
expressed as $(1-n^{R'}_+)n^N_+ - (1-n^N_+)n^{R'}_+$ can be interpreted as the 
probability of a $\rho$ meson scattering on a nucleon from the 
medium producing a resonance minus the process in which it
scatters from the resonance to produce a nucleon, 
the final states in both cases being Pauli-blocked. The corresponding processes
involving anti-baryons are included in the second term.

Let us now proceed to evaluate the integral over the momenta in the 
$NR^{-1}$ loop. The integral over $\cos\theta$ in 
$d^3p=-2\pi\sqrt{\omega^2_N-m^2_N}\omega_N d\omega_Nd(\cos\theta)$ is done 
using the $\delta-$function. Also the condition 
$|\cos\theta|\leq 1$ puts restriction on the range of 
integration over $\omega_N$.
A substantial simplification is obtained by changing the integration variable 
to $x$ using $\omega_N=\frac{S^2}{2q^2}(-q_0+|\vq|x)$ where 
$S^2=q^2-m^2_R+m^2_N$ to get
\be
{\rm Im}\Pi_B^\mn(q_0,\vq)=-\frac{S^2}{32\pi q^2}\int_{-W}^W dx\ L^\mn(x)
\{n_+(q_0+\omega_N)-n_+(\omega_N)+n_-(q_0+\omega_N)-n_-(\omega_N) \}
\label{piB_final}
\ee
where $W=\sqrt{1-4q^2m_N^2/S^4}$ and the factor $L^\mn$ in terms of the variable $x$ has the same form for
both diagrams in Fig.~1. This is shown in the appendix. 

The real part consists of principal value integrals which remain after removing the
imaginary part from (\ref{Pi_a}). Note that unlike the imaginary part, 
the real part of the self-energy at a given value of $q$
receives contribution from all the four terms.

Up to now we have been treating the baryon resonances $R$ in the narrow width 
approximation. It is indeed necessary to consider the width of the
unstable baryons in a realistic evaluation of the spectral function. 
For this, we follow the procedure (see
e.g.~\cite{SarkarNPA,Vijande})
of convoluting the self energy calculated in the narrow width 
approximation with the spectral function of the baryons. 
This approach has the advantage that the analytic structure of the self energy 
discussed above remains undisturbed.

\be
\Pi_B^\mn(q;m_R)= \frac{1}{N_R}\int^{m_R+2\Gm_R}_{m_R-2\Gm_R}dM\frac{1}{\pi} 
{\rm Im} \left[\frac{1}{M-m_R + \frac{i}{2}\Gm_R(M)} \right] \Pi_B^\mn(q;M) 
\ee
with $N_R=\displaystyle\int^{m_R+2\Gm_R}_{m_R-2\Gm_R}dM\frac{1}{\pi} {\rm Im} 
\left[\frac{1}{M-m_R + \frac{i}{2}\Gm_R(M)} \right]$ and 
$\Gm_R(M)=\Gm_{R\rightarrow N\pi} (M) + \Gm_{R\rightarrow N\rho} (M)$.
As a consequence of this convolution, the sharp ends of the regions of
non-zero imaginary part smoothly go to zero at a higher value of $M$
depending upon the width of the resonance. This is shown in
Fig.~\ref{impi_B} for the $N^*(1520)$ resonance. 

Nuclear medium at finite temperature is also substantially populated by mesons 
which modify 
the $\rho$ propagation in the medium 
in a non-trivial way. This has been studied~\cite{Ghosh1}
following the same procedure as described here for mesonic loop graphs with
one internal pion line and another meson line $h$ where $h=\pi,\om,h_1,a_1$
using interactions from chiral perturbation theory. 
Collecting the real and
imaginary parts, the self-energy from 
mesonic loops can be written as
\bea
&&\Pi^{\mn}_M(q_0,\vq)= \int \frac{d^3 \vec k}{(2\pi)^3}
\frac{1}{4\om_\pi\om_h}\left[-\frac{N^{\mn}_1 n_\pi - N^{\mn}_3 n_h}
{q_0 - \om_\pi + \om_h+i\epsilon(q_0)\eta}+\frac{N^{\mn}_2 n_\pi - N^{\mn}_4 n_h}
{q_0 + \om_{\pi} - \om_h+i\epsilon(q_0)\eta}
\right.\nonumber\\
&& \left.+ \frac{N^{\mn}_1 (1+n_\pi)  + N^{\mn}_4 n_h}{q_0 + \om_\pi - 
\om_h +i\epsilon(q_0)\eta}- \frac{N^{\mn}_2(1+ n_\pi) + N^{\mn}_3 n_h}
{q_0 + \om_\pi + \om_h +i\epsilon(q_0)\eta}\right]
\label{pi_meson}
\eea
where the  Bose distribution functions $n_\pi\equiv n(\omega_\pi)$ with
$\omega_\pi=\sqrt{\vk^2+m_\pi^2}$ and
$n_h\equiv n(\omega_h)$ with $\omega_h=\sqrt{(\vq-\vk)^2+m_h^2}$.
 $N^{\mn}_i(i=1,4)$ are the values of $N^{\mn}(k_0)$ for
$k_0=\om_\pi,-\om_\pi,q_0+\om_h,q_0-\om_h$ respectively and can 
be expressed in terms of the gauge-invariant tensors (\ref{ABC}).
The complete expressions are provided in the appendix. 
Using the procedure described above 
we have improved upon the
calculations in~\cite{Ghosh1} by including  the
width of the heavy mesons $a_1$ and $h_1$.
In this case we use a slightly different formula~\cite{Vijande}, 
\be
\Pi_M^\mn(q;m_h)= \frac{1}{N_h}\int^{(m_h+2\Gm_h)^2}_{(m_h-2\Gm_h)^2}dM^2\frac{1}{\pi} 
{\rm Im}\left[\frac{1}{M^2-m_h^2 + iM\Gm_h(M) } \right] \Pi_M^\mn(q;M) 
\ee
with $N_h=\displaystyle\int^{(m_h+2\Gm_h)^2}_{(m_h-2\Gm_h)^2}
dM^2\frac{1}{\pi} {\rm Im}\left[\frac{1}{M^2-m_h^2 + iM\Gm_h(M)} \right]$
and $\Gm_h(M)=\Gm_{h\rightarrow \rho\pi} (M)$.

The transverse and longitudinal components can then be obtained from the
self-energy tensors using the relations (\ref{pitpil}).

\section{Numerical Results}

\begin{figure}
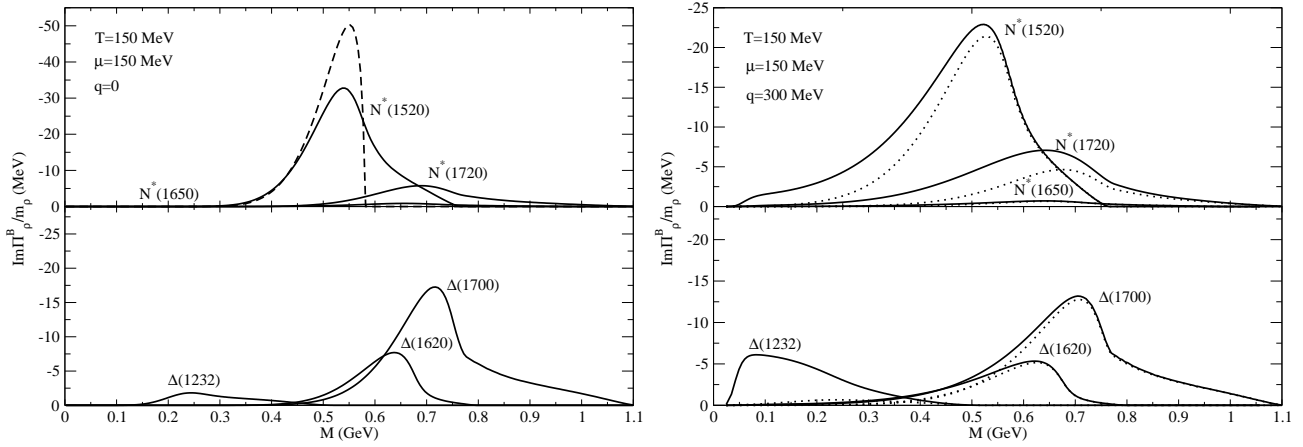

\includegraphics[scale=0.35]{im_qv_00.eps}
\includegraphics[scale=0.35]{im_qv_300.eps}
\caption{Imaginary part of $\rho$ meson self-energy  
showing the individual contributions for different $NR$ loops.
Left panel shows results for $\vq=0$ and the right panel shows the
transverse (solid) and longitudinal (dotted) parts for $\vq=300$ MeV. The 
long dashed line in the upper panel on the left shows
the imaginary part of the $NN^*(1520)$ loop evaluated in the narrow width approximation.}
\label{impi_B}
\end{figure}

\begin{figure}
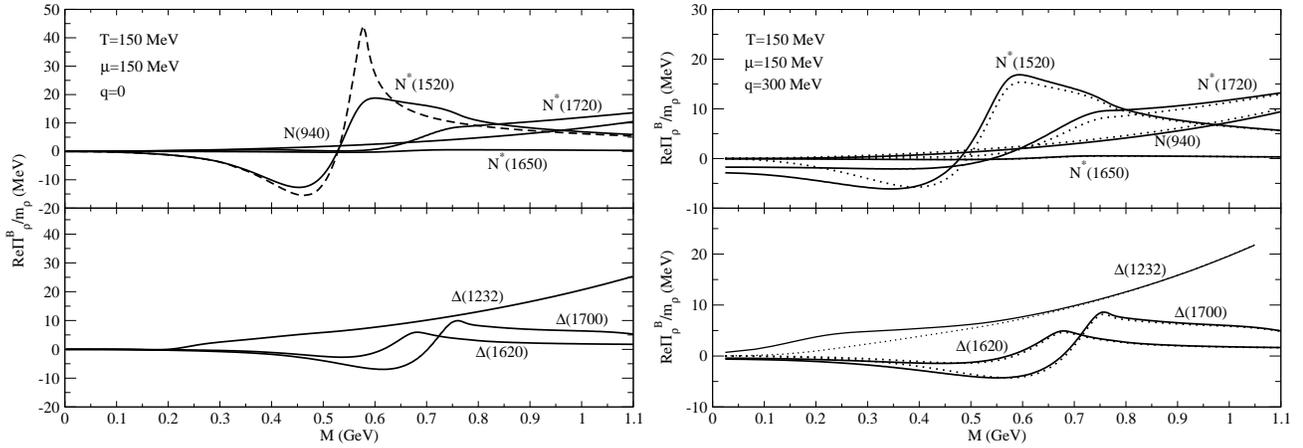

\includegraphics[scale=0.35]{re_qv_00.eps}
\includegraphics[scale=0.35]{re_qv_300.eps}
\caption{Same as Fig.~\ref{impi_B} for the real part}
% {Real part of $\rho$ meson self-energy  
%showing the individual contributions for different $NR$ loops.
%Left panel shows results for $\vq=0$ and the right panel shows the
%transverse and longitudinal parts for $\vq=300$ MeV.}
\label{repi_B}
\end{figure}

In this section we present the results of numerical evaluation beginning with the
imaginary part of the $\rho$ self-energy
as a function of the invariant mass $\sqrt{q^2}\equiv M$ for two values of the
three momentum. 
The Landau cut
contribution starts from $M=0$ for all values of the three momentum.
%for the loop graphs of Fig.~1.
Shown in Fig.~\ref{impi_B} left panel are the contributions from the individual
$NR$ loops for a $\rho$ meson at rest.
The $NN^*(1520)$ loop makes the most significant contribution followed by
the $N^*(1720)$ and $\De(1700)$. The right panel shows the 
corresponding results for $\vq=300$ MeV where the transverse and 
longitudinal components $\Pi_t$ and $q^2\Pi_l$ have been shown separately
by solid and dotted lines respectively.
(Note that for a $\rho$ meson at rest $\Pi_t=q_0^2\Pi_l$.)
The corresponding results for the thermal contribution to the real part are shown in Fig.~\ref{repi_B}.
The divergent vacuum contribution in this case
is assumed to renormalize the $\rho$ mass to its physical value. 
%Both the imaginary and real parts
%are found to decrease with increasing three momentum. 
Also shown by the long dashed lines in the left upper panels of Figs.~\ref{impi_B} and
\ref{repi_B} are the corresponding contributions to the real and
imaginary parts coming from the $NN^*(1520)$ loop computed in the narrow width
approximation.
\begin{figure}
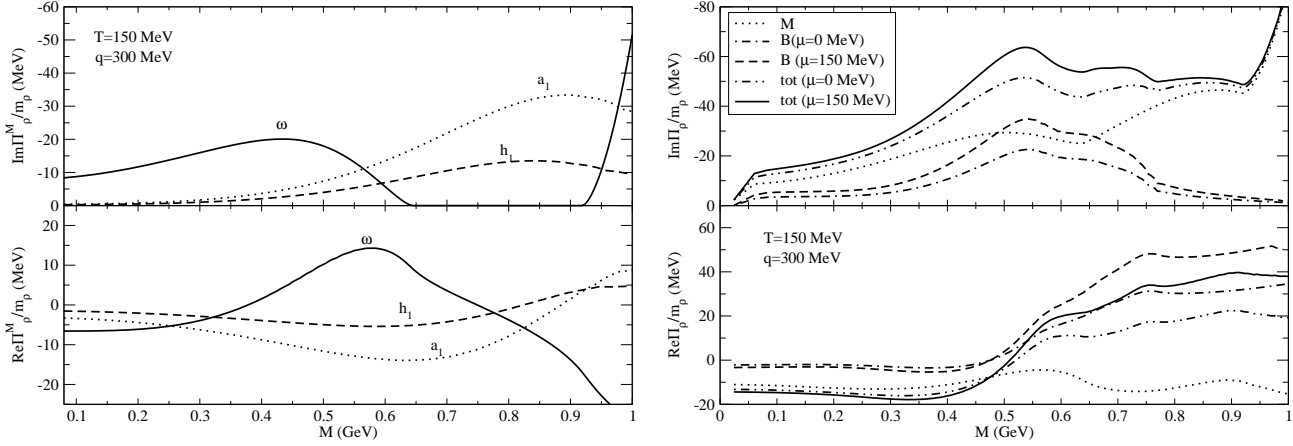

\includegraphics[scale=0.35]{im_re_M.eps}~~
\includegraphics[scale=0.35]{im_re_MBt.eps}
\caption{(Left) Imaginary and real parts of $\rho$ meson self-energy  
showing the individual contributions for different $\pi-h$ loops.
(Right) The total contribution from meson and baryon loops.}
\label{imrepi_MBt}
\end{figure}  

Next we show the results of the spin-averaged $\rho$ self-energy 
defined by 
\be
\Pi=\frac{1}{3}(2\Pi_t+q^2\Pi_l)~.
\ee
where the transverse and longitudinal components are obtained
using (\ref{pitpil}). In Fig.~\ref{imrepi_MBt} left panel the imaginary and real parts of $\pi-h$ loop
graphs are shown in the upper and lower panels respectively. 
The Landau and unitary cut contributions for the $\pi-\om$ loop are clearly
discernible though the contribution at the $\rho$ pole is dominated by the $a_1$.
On the right panel we plot the total contribution from the baryon and meson 
loops for two values of the baryonic chemical potential. The small positive
contribution from the baryon loops to the real part is partly compensated
by the negative contributions from the meson loops. The
substantial baryon contribution at vanishing baryonic chemical potential 
reflects the importance of anti-baryons.
\begin{figure}
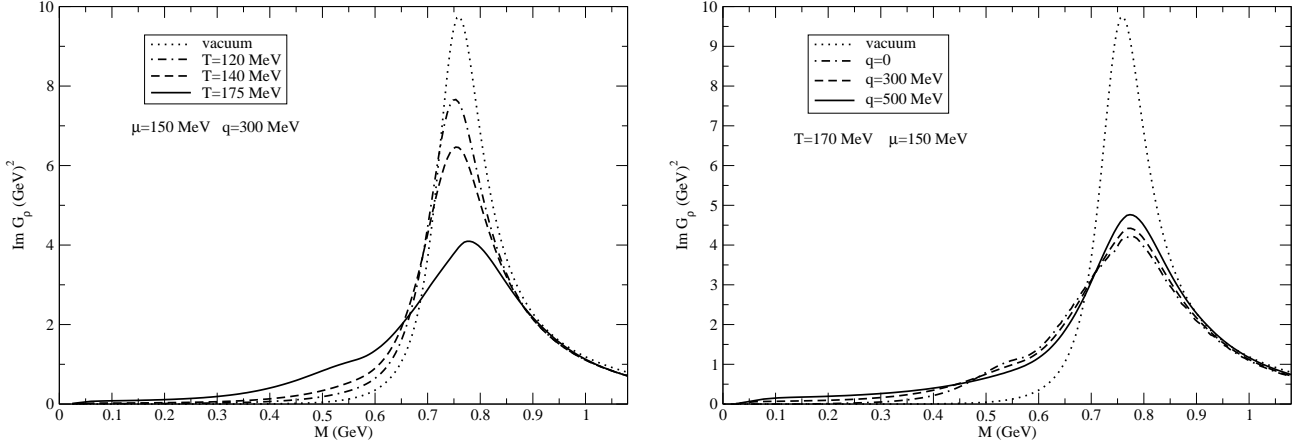

\includegraphics[scale=0.35]{spec_T.eps}~~~
\includegraphics[scale=0.35]{spec_qv.eps}
\caption{ The spectral function of the $\rho$ meson  
for (left) different values of the temperature $T$ and
(right) different values of the three-momentum $\vq$.}
\label{spec_T_qv}
\end{figure}

We now turn to the spin averaged spectral function given by
\be
\mathrm{Im}\,\og(q)=\frac{1}{3}(2\mathrm{Im}\,\og_t+q^2\mathrm{Im}\,\og_l)
\label{spavgG}
\ee
where
\be
\mathrm{Im}\,\og_{t,l}(q)=\frac{-\sum\iop_{t,l}^{\rm tot}}{(M^2-m_\rho^2-
(1,q^2)\sum\mathrm{Re}\,\overline\Pi_{t,l}^{\rm
tot})^2+\{(1,q^2)\sum\iop_{t,l}^{\rm tot}\}^2}~.
\ee
First, in Fig.\ref{spec_T_qv} left panel we plot the spectral function at fixed
values of the baryonic chemical potential and three-momentum for various
representative values of the temperature. 
We observe an increase of spectral strength at lower 
invariant masses resulting in broadening of the spectral function 
with increase in temperature. This is purely a Landau cut
contribution from the baryonic loops arising from the scattering of the $\rho$ from
baryons in the medium.
However, we do not observe much
variation with the three-momentum of the $\rho$ as seen from the figure on the
right panel.

\begin{figure}
\centerline{\includegraphics[scale=0.35]{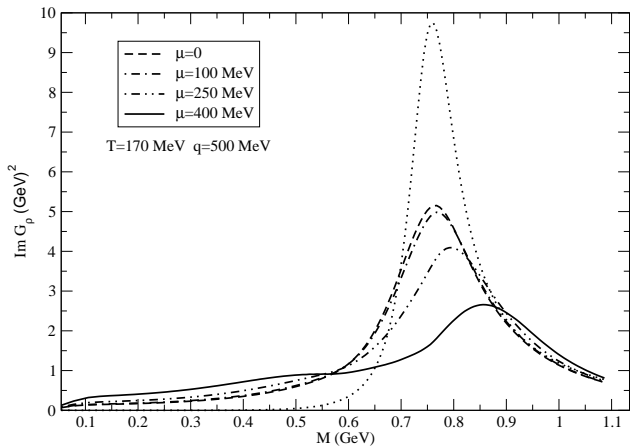}}
\caption{ The spectral function of the $\rho$ meson 
for different values of the baryonic chemical potential $\mu$}
\label{spec_mu}
\end{figure} 
We then plot in Fig.~\ref{spec_mu}, the spectral function for 
various values of the baryonic chemical potential for a fixed temperature.
For high values of $\mu$
we observe an almost flattened spectral density of the $\rho$.  

\begin{figure}
\centerline{\includegraphics[scale=0.35]{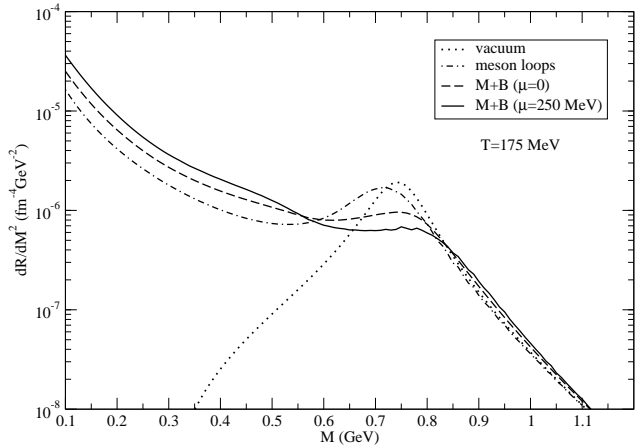}}
\caption{ The lepton pair emission rate at $T=175$ MeV with and without
baryon (B) loops in addition to the meson (M) loops}
\label{dilfig}
\end{figure}

Owing to differences in the various approaches to the evaluation of the $\rho$
spectral function followed in the literature, a direct numerical comparison  
with earlier results does not appear to be meaningful. These differences 
are at the level of the basic formulae arising from the type of couplings 
of the nucleon and the rho fields with the various baryon resonances considered
as well as in the form of the propagators used in the calculations. There
also exist differences at the level of formalism employed in the
evaluation of the $\rho$ self-energy.
We thus end this section by showing how the in-medium spectral function
of the $\rho$ is 
manifested in the dilepton emission rate.
This rate from thermalised 
hadronic matter is given by~\cite{Toimela}
\be
\frac{dR}{d^4q}=-\frac{\alpha^2}{3\pi^3 q^2}H(M^2)n_{BE}(q_0)\,g^\mn{\rm Im}
T_\mn(q_0,\vq)
\ee
where ${\rm Im}T_\mn$ is the imaginary part of the (retarded) two-point 
function of vector currents which can be obtained from (\ref{Tmed}) using
relations analogous to (\ref{def_diag}). The quantity $H(M^2)=(1+{2m_l^2}/{M^2})~
(1-4m_l^2/M^2)^{1/2}~$ is of the order of unity for electrons and will be
omitted henceforth.
In the low invariant mass ($M$) region, ${\rm Im}T_\mn$ is usually 
expressed as a sum over
the spectral densities of the vector mesons $\rho$, $\omega$ and
$\phi$~\cite{Annals}. This is however justified only in vacuum. 
The vector mesons $\rho$ and $\om$ can in general undergo mixing in the presence 
of matter which can lead to non-trivial modifications, 
for example, of the electromagnetic form factor
of the pion~\cite{Poda_JPG} and consequently the invariant mass distribution
of lepton pairs. In the following, we will consider only the
$\rho$ pole contribution which is known to play the most dominant role.
The dilepton rate in this case can be 
expressed in terms of the spin-averaged spectral function of the $\rho$ 
(\ref{spavgG}) getting,
\be
\frac{dR}{dM^2q_Tdq_Tdy}=\frac{\alpha^2}{\pi^2 M^2}n_{BE}(q_0)K_\rho{\rm Im}\ov G(q_0,\vq)
\ee
where we have used $T_\mn=K_\rho G_\mn$ as defined earlier. 
Integrating over the transverse momentum $q_T$ and rapidity $y$ of the 
electron
pairs we plot $dR/dM^2$ vs $M$ in Fig.~\ref{dilfig} for $T$=175 MeV. 
Because of the
kinematical factors multiplying the $\rho$ spectral function 
the broadening appears magnified in the dilepton
emission rate.
A significant enhancement is seen in the low mass lepton production rate
due to baryonic
loops over and above the mesonic ones shown by the dot-dashed line.
The substantial contribution from baryonic loops even for vanishing
chemical potential points to the important role played by antibaryons in thermal
equilibrium in systems created at RHIC and LHC energies. 

\section{Summary and Discussion}

We evaluate the $\rho$ self-energy to one loop in nuclear matter at finite
temperature and baryon density. Loop graphs involving the nucleon and 4-star
$N^*$ and $\De$ resonances up to spin 3/2 were calculated using gauge invariant
interactions in the framework of real time thermal field theory to obtain the
correct relativistic expressions for the $\rho$ self-energy. The singularities in
the complex energy plane were analysed and the imaginary part obtained from the
Landau cut contribution. Results for the real and imaginary parts at non-zero
three-momenta for various
values of temperature and baryonic chemical potential were shown for the
individual loop graphs. Adding the contributions from mesons obtained  
in the same formalism, the spectral function of the 
$\rho$ was observed to undergo a significant modification at and below 
the nominal rho mass which was seen to bring about a large enhancement of
lepton pair yield in this region.

It may be emphasised that the determination of the $\rho$ spectral
density at finite
temperature and baryon density by an explicit evaluation of loop
graphs using thermal field theoretic techniques such as performed here
is of relevance in view of precision data from experiments at
RHIC and LHC as well as from the
FAIR facility at GSI in future. But an actual comparison with data will 
involve a
space-time evolution of the static rates using a framework like
relativistic hydrodynamics.   
Efforts in this direction are in progress and
will be reported in due coarse.   

\section{Acknowledgement}

The authors gratefully acknowledge  discussions with S. Mallik
during the coarse of this work and to J. Alam for useful suggestions.

\section{Appendix}
\setcounter{equation}{0}
\renewcommand{\theequation}{A.\arabic{equation}}
\subsection{The real-time propagators and the self-energy matrix}

\begin{figure}
\centerline{\includegraphics{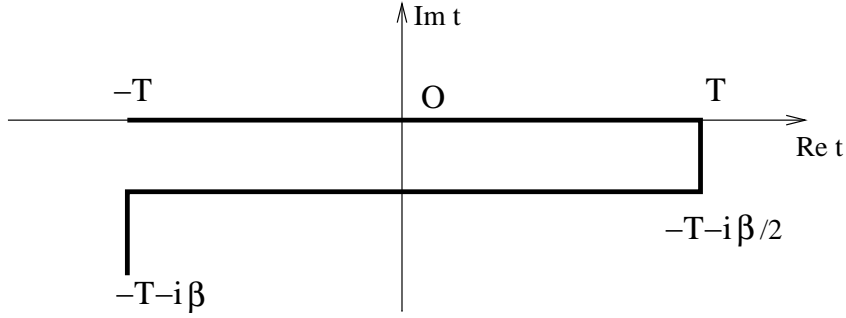}}
%\centerline{\psfig{figure=real_time.eps,height=2.5cm,width=8cm}}
\caption{Contour in the complex time plane for real time formalism}.
\label{contourfig}
\end{figure}

In the real time formulation of thermal field
theory a two-point function of local operators assumes a $2\times 2$ matrix
structure on account of the shape of the contour in the complex time plane.
Thus the current correlator $T^\mn$~\cite{SarkarIJP} as well as propagators for various
fields\cite{Kobes}
assume a matrix structure. For the contour shown in Fig.~\ref{contourfig} the
components of a free scalar propagator are as
\bea
&&D^{11}=-(D^{22})^*=\De(q)+2\pi in\de(q^2-m^2)\nonumber\\
&&D^{12}=D^{21}=2\pi i\sqrt{n(1+n)}\de(q^2-m^2)
\label{Dlam}
\eea
where $\De(q)$ is the Feynman propagator in vacuum,
\be
\De (q^2)=\frac{-1}{q^2 -m^2 +i\ep}~.
\ee
The thermal propagator may be diagonalised in the form 
\be
D^{ab}(q_0,\vec q)=U^{ac}(q_0)[{\rm diag}\{\De(q_0,\vq),-\De^*(q_0,\vq)\}]^{cd}
U^{db}(q_0)
\label{diag_D}
\ee
with the elements of the diagonalising matrix as
\[U^{11}=U^{22}=\sqrt{1+n},~~~ U^{12}=U^{21}=\sqrt{n}~.\]
Using the (transverse) vector propagator given by
\be
G^{(0)ab }_{\mn}(q)=\left(-g_{\mn}+\frac{q_\mu q_\nu}{q^2}\right)D^{ab}(q)
\ee
in the Dyson equation and the fact that $U$ diagonalises not
only the free propagator, but also the complete one~\cite{Kobes,Mallik_RT} it
turns out that the self-energy matrix $\Pi^{ab}_{\mn}$ is also diagonalisable by
$(U^{-1})^{ab}$, 
\be
\Pi^{ab}_{\mn}(q)=[U^{-1}(q_0)]^{ac}[{\rm
diag}\{\op_{\mn}(q),-\op_{\mn}^*(q)\}]^{cd}[U^{-1}(q_0)]^{db}~.
\ee
The relations 
\bea
{\rm Re}\,\op_{\mn}&=&{\rm Re}\,\Pi_{\mn}^{11}\nonumber\\
\iop_{\mn}&=&\epsilon(q_0)\tanh(\beta q_0/2){\rm Im}\,\Pi_{\mn}^{11}
\eea
trivially follow showing that the thermal matrices are actually given by a {\em single} 
analytic function which essentially coincides with the corresponding 
result from the imaginary time formulation.

\subsection{The factor $L^{\mn}$ in baryonic loops}

The factor $L^{\mn}$ which appears in the loop integrals is given by
\be
L^{\mn}(p,q)=I_F \left(\frac{fF(q)}{m_\rho}\right)^2 [\alpha A^{\mn}+\beta
B^{\mn}+\gm C^{\mn}]
\ee
where $A^{\mn},\,B^{\mn}$ and $C^{\mn}$ are the transverse tensors defined in
(\ref{ABC}). The coefficient factors $\alpha,\,\beta,\,\gm$ are given in Table~1 where
\bea
\alpha_{1/2+}&=&4q^2(p^2-m_N m_R+p\cdotp q)\nn\\
\alpha_{1/2-}&=&\alpha_{1/2+}(m_N\to -m_N)\nn\\
\beta_{1/2}&=&8\nn\\
\gm_{1/2}&=&0\nn\\
\alpha_{3/2+}&=&\frac{2q^2}{3m_R^2}[p^2(p^2-3q^2)+p\cdot q(3p^2+q^2)\nn\\
&&+3m_R^2(p^2+2m_Nm_R+p\cdot q)\nn\\
&&-2m_Nm_R(p^2+q^2-2p\cdot q)]\nn\\
\alpha_{3/2-}&=&\alpha_{3/2+}(m_N\to -m_N)\nn\\
\beta_{3/2}&=&4(1 +{p^2}/{3m_R^2})\nonumber\\
\gamma_{3/2}&=&-{4}/{3m_R^2}~.
\eea

\begin{table}%[h]
\begin{center}
%\label{tab}
\begin{tabular}{|c|c|c|c|c|c|}
\hline
& & & & &\\
$R$ & $J^P$ & $f$ & $\alpha$ & $\beta$ & $\gm$ \\
& & & & &\\
\hline
& & & & &\\
$N(940)$ & $\frac{1}{2}^+$ & 7.7 & $\frac{1}{2}\alpha_{1/2+}$ & $\frac{1}{2}\beta_{1/2}$ &
$\frac{1}{2}\gm_{1/2}$ \\
& & & & &\\
$N^*(1520)$ & $\frac{3}{2}^-$ & 7.0 & $\alpha_{3/2-}$ & $\beta_{3/2}$ & $\gm_{3/2}$ \\
& & & & &\\
$N^*(1650)$ & $\frac{1}{2}^-$ & 0.9 & $\alpha_{1/2-}$ & $\beta_{1/2}$ & $\gm_{1/2}$ \\
& & & & &\\
$N^*(1720)$ & $\frac{3}{2}^+$ & 7.0 & $\alpha_{3/2+}$ & $\beta_{3/2}$ & $\gm_{3/2}$ \\
& & & & &\\
$\Delta(1232)$ & $\frac{3}{2}^+$ & 10.5 & $\alpha_{3/2+}$ & $\beta_{3/2}$ & $\gm_{3/2}$ \\
& & & & &\\
$\Delta(1620)$ & $\frac{1}{2}^-$ & 2.7 & $\alpha_{1/2-}$ & $\beta_{1/2}$ & $\gm_{1/2}$ \\
& & & & &\\
$\Delta(1700)$ & $\frac{3}{2}^-$ & 5.0 & $\alpha_{3/2-}$ & $\beta_{3/2}$ & $\gm_{3/2}$ \\
& & & & &\\
\hline
\end{tabular}
\caption{Table showing the coefficients $\al$, $\beta$ and $\gm$ for loops 
containing the  various resonances considered.}
\end{center}
\end{table}

The corresponding expressions of $\alpha,\beta$ and $\gamma$ for the second 
graph of Fig.~1 can be obtained by replacing $q\to-q$ as indicated in 
(\ref{diag1a-b}).
The factor 1/2 in front of the coefficients for the $NN$ loop is put
to prevent a double counting of this contribution. 
The coupling constants $f$ have been
obtained as in~\cite{Peters,Post2}. The values are also in reasonable agreement
as seen in Table~1.

%\subsection{Simplification of the factor $L_{\mn}$ in the imaginary part}

A considerable simplification in the imaginary part
can be achieved in the sum of the two diagrams
in Fig.~1 by a change of variables. 
Note that only the factor $p\cdot q$ which appears with various powers in the tensors $A^\mn,\,
B^\mn$ and $C^\mn$ and in the factors $\alpha$
takes on different values in the various terms in the expression for the imaginary part of 
self energy. We recall for convenience the imaginary part coming from 
first diagram in Fig.~1,
\bea
&& {\rm Im} \ov\Pi^{(1)}(q_0,\vq)=-\pi \int\frac{d^3\vec p}{(2\pi)^3 4\om_N\om_R}\times  \nonumber\\
&&[L_1(q)\{(1-n^N_+-n^R_-)\de(q_0-\om_N-\om_R)
+(n_+^N-n_+^R)\de(q_0-\om_N+\om_R)\}\nonumber\\
&& + L_2(q)\{(n_-^R-n_-^N)\de(q_0+\om_N-\om_R)
-(1-n_-^N-n_+^R)\de(q_0+\om_N+\om_R)\}]
\label{impi1}
\eea 
where we have dropped the Lorentz indices for brevity. For the first two 
terms $p_0=\omega_N$ and on integration over the angle
using either of the two delta functions one has 
$\opq=-\frac{1}{2}(S^2-2q_0\omega_N)$ with $S^2=q^2-m^2_R+m^2_N$. One thus gets
$p\cdot q=p_0q_0-\opq=\frac{S^2}{2}$. Changing variable to $x$ 
using $\omega_N=\frac{S^2}{2q^2}(q_0+|\vq|x) $ one has
\bea
&& A_\mu^\mu =-3,~~~~ A_{00}=\frac{|\vq|^2}{q^2}\\
&& B_\mu^\mu =m_\pi^2 q^2 +\frac{S^4}{2},~~~~ B_{00}=-\frac{|\vq|^2S^4}{4q^2}(1-x^2)\\
&&  C_\mu^\mu =q^2(m_\pi^2 q^2 -\frac{S^4}{4}),~~~~
C_{00}=\frac{|\vq|^2S^4}{4}x^2~.
\label{abc00}
\eea
For the last two terms, $\opq=-\frac{1}{2}(S^2+2q_0\omega_N)$ and 
$p_0=-\omega_N$ so that here too $p\cdot q=\frac{S^2}{2} $.
Defining $x$ in this case through $\omega_N=
\frac{S^2}{2q^2}(-q_0+|\vq|x)$ one recovers the
same expression for the tensor component as in the first two terms.

The corresponding expression for the imaginary part from the second graph of Fig.~1 
is obtained by changing $q\rightarrow -q$ in the expression for 
${\rm Im}\Pi$  given by (\ref{impi1}) getting
\bea
&& {\rm Im} \ov\Pi^{(2)}(q_0,\vq)=-\pi \int\frac{d^3\vec p}{(2\pi)^3 4\om_N\om_{R'}}\times  \nonumber\\
&&[L_2(-q)\{(1-n^N_--n^{R'}_+)\de(q_0-\om_N-\om_{R'})
+(n_-^N-n_-^{R'})\de(q_0-\om_N+\om_{R'})\}\nonumber\\
&& + L_1(-q)\{(n_+^{R'}-n_+^N)\de(q_0+\om_N-\om_{R'})
-(1-n_+^N-n_-^{R'})\de(q_0+\om_N+\om_{R'})\}]
\eea 
where $\omega_{R'}=\sqrt{(\vp+\vq)^2+m_{R'}^2}$.
In this case, for the first two terms $\opq=\frac{1}{2}(S^2-2q_0\omega_N)$ 
which along with 
$p_0=-\omega_N$ gives $p\cdot q=-\frac{S^2}{2} $.
The same is obtained for the last two terms for which 
$\opq=\frac{1}{2}(S^2+2q_0\omega_N)$ and $p_0=\omega_N$. 
Making a change of variables
as before, identical values of the tensor components are obtained as in
(\ref{abc00}) 
as a consequence of the fact that the gauge invariant tensors
$A_{\mn},\, B_{\mn}$ and $ C_{\mn}$ are even under $q\rightarrow -q$.
The coefficients $\alpha,\, \beta$ and $\gamma$ also remain unchanged in the two
diagrams 
the sign of $p\cdot q$ in the expressions remaining the same under the 
combined effect of $q\rightarrow -q$ and a reversal in its magnitude
($\frac{S^2}{2}\rightarrow -\frac{S^2}{2}$). The value of $L(x)$ thus comes out 
to be the same for all cases and we end up with the final expression 
given by (\ref{piB_final}).

%For the real part however, this simplification does not occur and one has to
%explicitly change $q\rightarrow -q$ in the coefficients
%$\alpha, \beta, \gamma$ for the second diagram Fig~1.

\subsection{Expressions for $N_\mn$ for mesonic loops}

The expressions for $N_\mn$ appearing in the $\rho$ self-energy (\ref{pi_meson}) 
for $\pi-h$ loops have been obtained in~\cite{Ghosh1} and are given below,
\bea
N_{\mn}^{(\pi)}(q,k)&=&\left(\frac{2G_\rho}{m_\rho \F^2}\right)^2
C_{\mn}\nonumber\\
N_{\mn}^{(\om)}(q,k)&=&-4\left(\frac{g_1}{\F}\right)^2(B_{\mn}+q^2k^2
A_{\mn})\nonumber\\
N_{\mn}^{(h_1)}(q,k)&=&-\left(\frac{g_2}{\F}\right)^2(B_{\mn}-\frac{1}{m_{h_1}^2}
C_{\mn})\nonumber\\
N_{\mn}^{(a_1)}(q,k)&=&-2\left(\frac{g_3}{\F}\right)^2(B_{\mn}-\frac{1}{m_{a_1}^2}
C_{\mn})
\eea
where the constants $G_\rho=69$ MeV, $\F=93$ MeV, $g_1=0.87,\, g_2=1.0$ and $g_3=1.1$. 
These can be simplified as shown above and finally expressed in terms of
the tensor components defined in (\ref{abc00}). As in the case of baryon loops
the factors $N_{\mn}$ have also been multiplied by the square of
the monopole form factor $F(q)$ defined earlier.

\end{document}